  \documentclass{aastex}
  \usepackage{emulateapj5}
\newcommand{\eg}{e.g., } 
\newcommand{\ie}{i.e., } 

\begin{document}

\title
{
 Spectral Consequences of Deviation from Spherical Composition Symmetry
 in Type Ia Supernovae
}


\author
{
 R. C. Thomas\altaffilmark{1}, Daniel Kasen\altaffilmark{2}, 
 David Branch\altaffilmark{1,3} and E. Baron\altaffilmark{1,3}
}

\altaffiltext{1}
{ 
 University of Oklahoma, Department of Physics and Astronomy,
 Norman, OK 73019, USA, thomas,branch,baron@mail.nhn.ou.edu
}

\altaffiltext{2}
{
 University of California, Berkeley, Department of Physics,
 Berkeley, CA 94720, USA, dnkasen@panisse.lbl.gov
}

\altaffiltext{3}
{
 Lawrence Berkeley Laboratory, Berkeley, CA 94720
}

\setcounter{footnote}{0}


\begin{abstract} \vskip .2 truein 

We investigate the prospects for constraining the maximum scale of 
clumping in composition that is consistent with observed Type Ia
supernova flux spectra.  Synthetic spectra generated without purely
spherical composition symmetry indicate that gross asymmetries make
prominent changes to absorption features.  Motivated by this,
we consider the case of a single unblended line forming in an 
atmosphere with perturbations of different scales and spatial distributions.
Perturbations of about 1\% of the area of the photodisk simply weaken
the absorption feature by the same amount independent of the line of
sight.  Conversely, perturbations of about 10\% of the area of the 
photodisk introduce variation in the absorption depth which does
depend on the line of sight.  Thus, 1\% photodisk area perturbations
may be consistent with observed profile homogeneity but 10\% 
photodisk area perturbations can not.  Based on this, we suggest that
the absence of significant variation in the depths of Si II $\lambda$ 6355
absorption features in normal Type Ia spectra near maximum light indicates
that any composition perturbations in these events are quite small.
This also constrains future three-dimensional explosion models to 
produce ejecta profiles with only small scale inhomogeneities.

\end{abstract}



\section{Introduction} 
\label{sec-intro} 

Three-dimensional calculations of white dwarf deflagrations suggest
significant deviation from spherical composition symmetry in the
resulting Type Ia supernova (SN) envelopes \citep{K00,H00}.  Yet the 
well-documented spectral homogeneity of normal events tightly
constrains the model output spectra.  This inspires us to consider a
very general question relevant to three-dimensional calculations of
white dwarf explosions:  How much and what kinds of deviation from
spherical composition symmetry might still produce morphologically
plausible and homogeneous spectra from all lines of sight?

The present Letter addresses this question using some simple arguments.
In \S 2, using a modified version of the direct analysis code SYNOW,
we generate some sample spectra to demonstrate artifacts caused by
deviation from spherical composition symmetry.  Motivated by these
results, we pursue in \S 3 a possible means of constraining deviation
from spherical composition symmetry by considering the formation of a single
unblended line in an envelope with a simple composition perturbation.
The small scatter in the depths of Si II absorption features in normal Type
Ia spectra near maximum light may indicate that any composition 
perturbations present exist only on small scales.


\section{Clumpy SN Spectra}
\label{sec-spectra}

The fast, parameterized, SN spectrum synthesis code SYNOW has been used
for ``direct'' analysis \cite{F97} of several SNe of varying types [\eg 
Type Ia SN 1994D \cite{H99}, Type Ib SN 1999dn \cite{D00}, Type Ic SN
1994I \cite{M99}, Type II SN 1999em \cite{B00}].  The goal of direct
analysis is to establish line identifications and intervals of
ejection velocity within which the presence of lines of various ions
is detected, without adopting any particular hydrodynamical model.
Composition and velocity constraints obtained from SYNOW then can
provide guidance to those who compute hydrodynamical explosion models
and to those who carry out computationally intensive NLTE spectrum
modeling.  A complete description of the workings of SYNOW can be
found in Fisher (2000).  Here we present only the background
necessary to understand the modifications made to allow SYNOW to
produce spectra from model envelopes without spherical composition
symmetry.

\subsection{SYNOW and ClumpySYN}

In its simplest form, SYNOW uses spherical composition symmetry and
the assumption of resonance scattering for line formation.  Line formation
takes place in an envelope surrounding a sharp photosphere, a source
of continuum radiation.  SYNOW treats line formation in the Sobolev
approximation, which is a good one for analysis purposes.  The
profile of a line is determined by the adopted radial distribution of 
the line optical depth and the line source function.

The geometric algorithm that SYNOW uses to calculate the line source
function at a given point in the envelope requires some explanation.
Consider an atmosphere like that depicted in Figure 1.  The shaded
region represents the sharp photosphere, and we draw an imaginary
boundary at $v_{max}$ where the optical depth in all lines we are
considering drops to a negligible amount.  Our goal is to calculate
the source function in some line with wavelength $\lambda$ at the 
point P.

Adopting the simple explosion velocity law $v \propto r$ implies that
surfaces of constant velocity with respect to the point P in the
atmosphere are spheres centered on that point.  In an envelope with
spherical composition symmetry, points physically relevant for
computing the line source function at P are those on such
common point velocity spheres which are within $v_{max}$ of the
explosion center and are not occulted by the photosphere.  Two such
surfaces are labelled A and B in the figure.  The cone C denotes the
boundary of an occultation region produced by the photosphere.

Now consider two other lines with wavelengths $\lambda_A$ and $\lambda_B$ 
such that $\lambda = \lambda_A (1 + \Delta v_A/c) = \lambda_B (1 + \Delta
v_B/c)$ where $\Delta v_A$ and $\Delta v_B$ are the velocity radii of
spheres A and B respectively.  Photons that redshift with respect to
the matter into resonance with the line $\lambda_B$ along the surface
B scatter isotropically.  Some of these are
redirected towards P.  These photons redshift further until they come
into resonance with the line $\lambda_A$ at surface A.  Here they can
be scattered away from P or proceed unhindered to redshift into
resonance with the line $\lambda$ at P.  Similarly, photons emitted
from the photosphere that redshift into resonance with the line
$\lambda_A$ on the sphere A can be scattered towards or away from P.
By computing the intensity arriving at P from many surfaces and the
photosphere, the source function in the line $\lambda$ at P can be
built up.  Clearly, computing the source function as a function of
radius is the most intensive part of a SYNOW calculation as the
program considers a long list of lines.  Even so, an entire spectrum
covering many lines can be produced in minutes.

One special case that occurs when calculating the source function is
when no shorter wavelength lines are close enough to $\lambda$ in
wavelength to be scattered from a corresponding surface within the
sphere $v_{max}$.  In such a case, the line is {\it unblended} and the
source function consists only of the photospheric intensity times the
geometric dilution factor of the photosphere.

When spherical composition symmetry is broken, a particular scheme
must be chosen to parameterize the line optical depths on a
three-dimensional grid instead of along a radius.  We call the 
code we use in this case ``ClumpySYN,'' and choose to group
together regions in the atmosphere in spherical ``clumps.''  We can
confine species to the clumps by setting the corresponding line
optical depths to zero everywhere else.  Conversely, we can exclude
species from clumps by setting corresponding line optical depths to
zero within clumps while assigning nonzero values elsewhere.  We still 
adopt a particular radial profile ($\tau \propto \exp(-v/v_e)$) to 
determine the value of the line optical depth at a point if it is
allowed to be nonzero there.

Unfortunately, distributing line optical depth in a way that breaks
spherical symmetry introduces a new wrinkle.  Figure 2 illustrates
that when optical depth is distributed in an arbitrary fashion,
different regions of the surfaces A and B contribute to the intensity
at P in an angular dependent way.  The above geometric algorithm for
determining the source function everywhere can either be generalized
to handle such arbitrary optical depth distributions (instead of a
simple function of radius only) or we can ignore line blending.  When
we compare two spectra generated with the same spherically symmetric
composition but with and without line blending (or multiple
scattering), we find that the differences between them are
minor.  Hence, in ClumpySYN we choose to neglect 
multiple scattering when calculating the source function.  Now the
source function in each line simply equals the photospheric intensity 
times the geometric dilution factor of the photosphere.

A new geometric algorithm could be developed to compute the source
function including multiple scattering, but since SYNOW makes 
several assumptions already (\ie pure presonance scattering source
function, a sharp photosphere) such an improvement seems only
marginally profitable.  This is especially clear in light of the
similarity of spherically symmetric synthetic spectra with and without
multiple scattering.

\subsection{Sample ClumpySYN Spectra}

An example of a clump configuration is shown in Figure 3.  Here, we
place an upper velocity boundary at 25000 km s$^{-1}$.  We allow the
clumps to overlap and have radii in velocity space between 5000 and 
6000 km s$^{-1}$. The fraction of the volume taken up by clumps in the
envelope (between the photosphere at 11000 km s$^{-1}$ and the upper
boundary) in Figure 3 is about 66\%.  Several such models were
generated with different volume filling factors, and model output
spectra were computed from several lines of sight with ClumpySYN.

Motivated by Figure 12 of Khokhlov (2000), we partition the clumps
into two parts.  In the inner part of the clumps, we place Fe II ions.
In the outer part, we place intermediate mass ions (Si II, Ca II, S II).  
Outside the clumps, we place O I.  The choice of these particular ions 
is motivated by a maximum light fit of SN 1994D \cite{H99}.

Figure 4 presents synthetic spectra from models with different envelope
filling factors.  Each graph displays four spectra, one each of four 
different lines of sight spaced 90 degrees apart about the equator.  We
note that higher volume filling factors cover the photosphere more
effectively and the spectra are quite similar along all lines of
sight.  At lower filling factors, perspective-dependent spectral
diversity begins to creep in, particularly in absorption features.


\section{The Threshold Clumping Scale}
\label{sec-threshold}

Previous studies of SN spectra from geometries departing from
spherical symmetry are polarization investigations 
\citep[\eg][]{H01,W97} and some axisymmetric configuration line
profiles \cite{JB90}.  These consider ellipsoidal deformations of the
envelope.  Here we are concerned with less global perturbations in
composition.

To illustrate clearly the effects of composition clumping on spectra,
we consider the formation of a single unblended line in an expanding
SN envelope.  The standard assumptions of SYNOW we also apply here:
homologous expansion and resonance scattering.  In this simple
exercise, we characterize the degree of clumping with two parameters.
The first parameter governs the cross sectional area of
an individual clump relative to the area of the photodisk (the
photosphere's cross section along a line of sight).  The second 
controls the spatial
deployment of the clumps, their sparsity or density.  Within
clumps, line optical depth is defined relative to its value at the
photosphere ($\tau \propto \tau_{phot} \exp(-v/v_e)$).

With the line source function known everywhere and $\tau$ assigned to
clumps, computing the line profile is simple.  We integrate over each
surface of common velocity relative to an observer at infinity to
obtain the flux as a function of velocity (or observer frame
wavelength).  In homologous flow, these surfaces are planes
perpendicular to the line of sight.

We define the ``photodisk covering factor'' for a plane of common velocity
as the ratio of the area in clumps intersecting that plane which are
contained in the projected photodisk area to the total photodisk area.
In a spherically symmetric composition, $\tau$
is greatest at the photosphere so the deepest part of the absorption
feature corresponds to the common velocity plane tangent to the
photosphere (unless the line is very strong).  Clearly, decreasing the
covering factor in this plane weakens the absorption minimum relative
to the ``totally covered photodisk'' minimum.  One could compensate
for this weakening by increasing $\tau_{phot}$, but there is a limit to
how effective this can be; even as
$\tau_{phot}$ is increased to infinity, the minimum flux will saturate 
because the
uncovered portion of the photodisk shines through.  This implies that
for a given observed absorption depth fit by some choice of
$\tau_{phot}$ assuming total photodisk coverage, there is some minimum
covering factor determined by $\tau_{phot}\rightarrow\infty$ that could 
fit the same line.

If the average clump is much smaller in cross section than the
photodisk area, the covering factor will tend to be the same value
from all lines of sight, inducing only limited perspective-dependent
absorption depth diversity.  On the other hand, if the average clump
has a large cross section comparable to the photodisk area, observers
with different lines of sight will measure different line profiles.
The covering factor as the line of sight is shifted will vary
enormously.

To illustrate these facts, $\tau$ is assigned to a three
dimensional Cartesian grid using the following ``white noise''
prescription.  The grid is partitioned into regular cubical cells with
$s$ grid points along each edge.  The list of cubical cells is then
traversed 
and at each cell a uniform random deviate is chosen.  If the deviate
is less than a prescribed value $f_c$, $\tau$ is assigned to each
grid point contained within the cell according to the exponential
$\tau$ profile, otherwise $\tau$ is set to zero at all points in the
cell.  In the end, for $N$ cells, about $f_c N$ will be cubical clumps
and the remainder will be empty.  We choose to compute line profiles
for six observers situated at infinity such that the lines of sight
are perpendicular to cell faces so the value of $f_c$ will roughly equal
the fraction of grid points within clumps on each common velocity plane.
The cubical shape of the clumps causes the line profiles to be
jagged, but other choices will just smooth out the line profiles and
not change the general result, since the photodisk covering factor is the
important quantity.

In Figures 5a and 5c, line profiles for two values of $s$ (8 and 32
grid points respectively) are shown.  In each figure, five values of
$f_c$ are shown, and six different viewpoints of each realization are
superimposed.  The photodisk radius is 50 grid points and the optical
depth at the photosphere is set to 7.5.  The optical depth falls off
with radius exponentially, with an e-folding length of 15 grid points.
In Figure 5a, the ratio of each clump cross section to the photodisk
area is small, about 1/120.  Hence, for all values of $f_c$, the line
profiles are quite similar from all lines of sight.  However, in
Figure 5c, each clump cross section is 1/8 the area of the photodisk
for the six observers, so changing perspective radically alters the
actual fraction of the photodisk covered.  This introduces 
perspective-dependent line profile diversity.

Figures 5b and 5d are plots of absorption feature depth versus $f_c$.
For $s=8$ (5b), the depths measured by each observer are close
together while for $s=32$ (5d), the depths are scattered markedly.  In
both figures, a vertical line indicates the lowest $f_c$ value
where $\tau_{phot}$ could be increased to match a $f_c=1$ observation.
A collection of spectra which exhibit significant absorption depths
consistent across the observed sample indicate that any clumps
present must be small and densely deployed, \ie a consistent, deep
line implies that the photodisk is nearly covered by optical depth
from all lines of sight.

In normal Type Ia SNe, one possible spectrum feature that might help
constrain the actual amount of clumping is the distinctive  Si II
doublet ($\lambda \lambda$ 6347, 6371 \AA) near maximum light.  We
studied a sample of good quality maximum light spectra obtained over
the last two decades and found that measured Si II depth does not vary
appreciably from event to event.  A few examples in the
wavelength range of interest are shown in Figure 6.  The average value
of the absorption depth is about 0.67 relative to the continuum, with
a standard deviation of 0.06.  If we interpret these observations as
instances of the same class of event observed from different lines of
sight, the lack of substantial scatter implies that the composition
clump scale of Si in normal Type Ia SNe is similar to that used to create line
profiles in Figure 5a.  Any clumps of Si present must be smaller than
the detection threshold for clumping, which we have illustrated to be
set by clumps much smaller than 1/10 the area of the photodisk.


\section{Conclusion}
\label{sec-conclusion}

By effecting minor modifications to the direct analysis code SYNOW, we
have produced ClumpySYN, a code that can generate spectra from
compositions without spherical symmetry.  Based on a simple
parameterization, we suggest that the important factor for spectra
from events lacking a spherical composition is the fraction of the
photosphere covered by clumps.  If the clumps are large compared to
the size of the photosphere, inhomogeneity in unblended line
absorption becomes manifest.  Below a threshold scale, these clumps
only weaken absorption along different lines of sight by the same
amount.  Absorption strength accounted for by spherically symmetric
compositions can be recovered in clumpy models by increasing optical
depth, but only to an extent.

In addition, we suggest that the robustness of absorption depth
measurements in the Si II feature in normal Type Ia SNe implies that any 
perturbations in
composition away from spherical symmetry are smaller than the
threshold scale.  From this, one might conclude that if a general
characteristic of deflagration models is the formation of large clumps
of Si and Fe, then normal Type Ia SNe are not the result of
deflagrations.  More conservatively, we suggest that explosion
models must avoid generating large-scale bubbles or clumps in
composition to recover normal Type Ia SN spectral homogeneity.


\acknowledgements

The authors wish to thank members of the University of Oklahoma
Supernova Research Group, David Jeffery and the anonymous referee
for comments and suggestions.  Partial
support for this work was provided by NSF grants AST-9731450
and 9986965, and NASA grant NAG 5-3505.



\clearpage

\figcaption[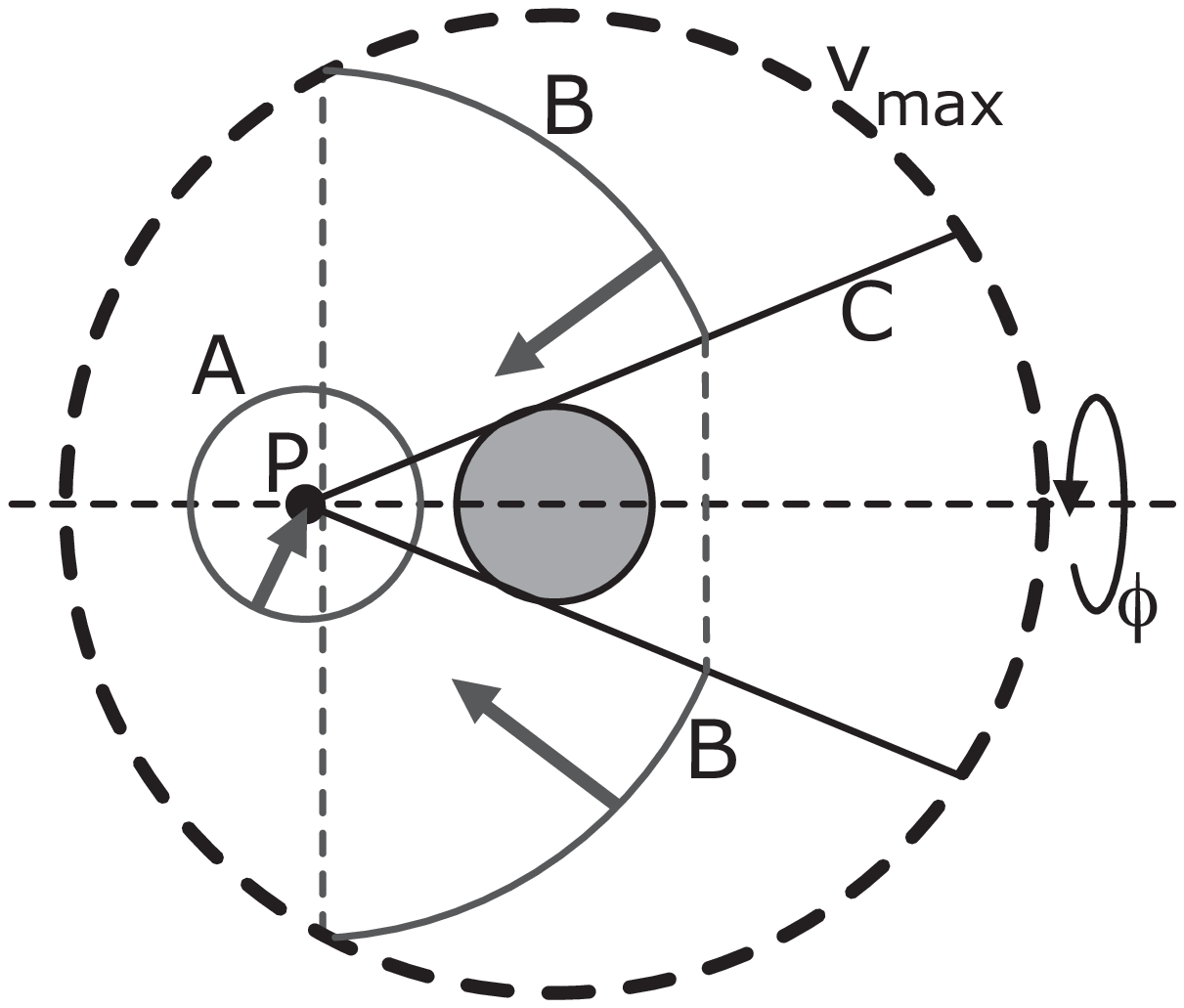]
{
 SYNOW computes the line source function at P by considering photons
 scattered there from cylindrically symmetric common point velocity 
 surfaces A and B.  The horizontal dashed line is the axis of symmetry for the 
 common point velocity surfaces as observed from P.
 \label{fig1}
}

\figcaption[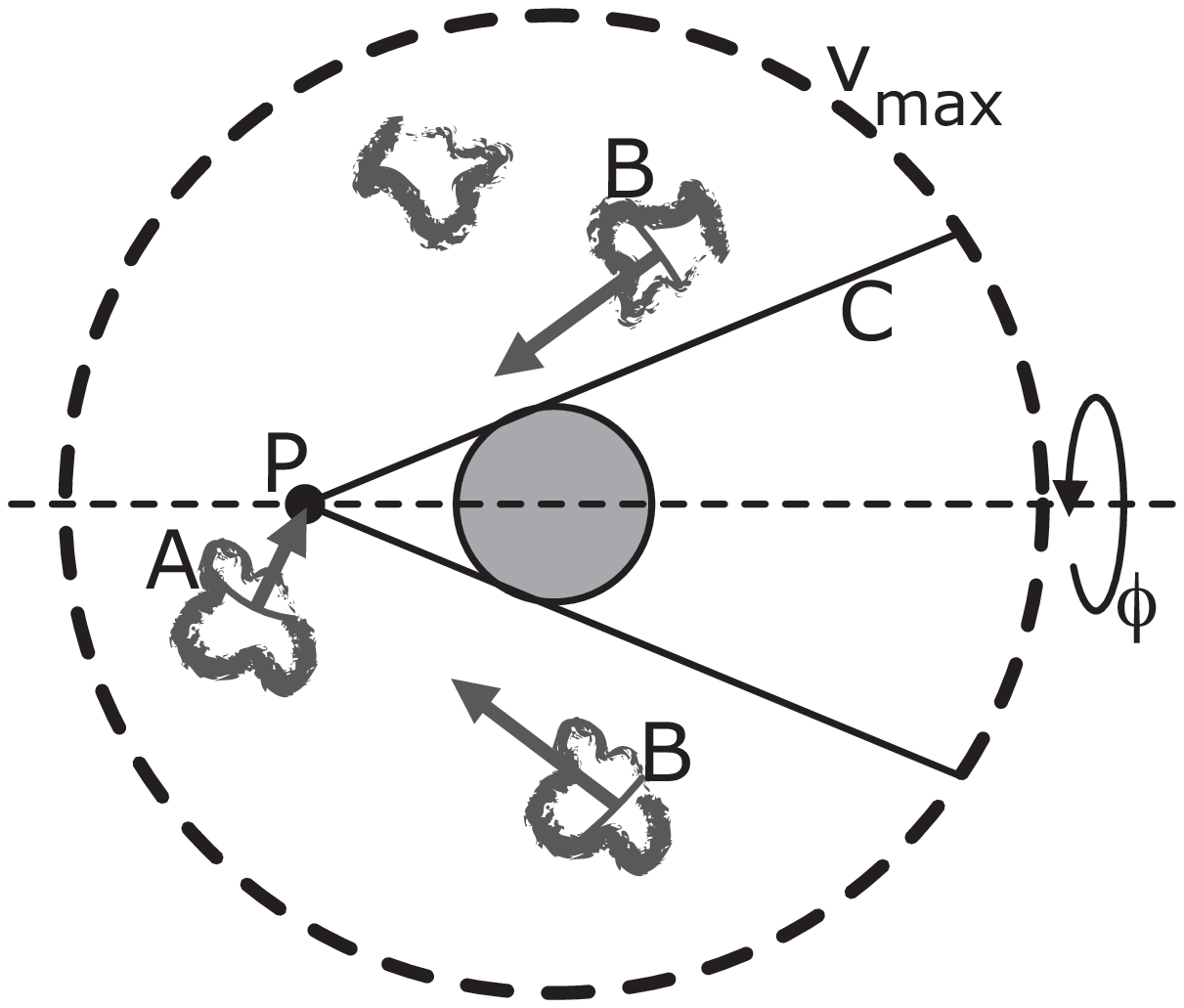]
{
 In cases where optical depth is not distributed in a spherically
 symmetric way, photons scatter toward P in an angular dependent way.
 \label{fig2}
}

\figcaption[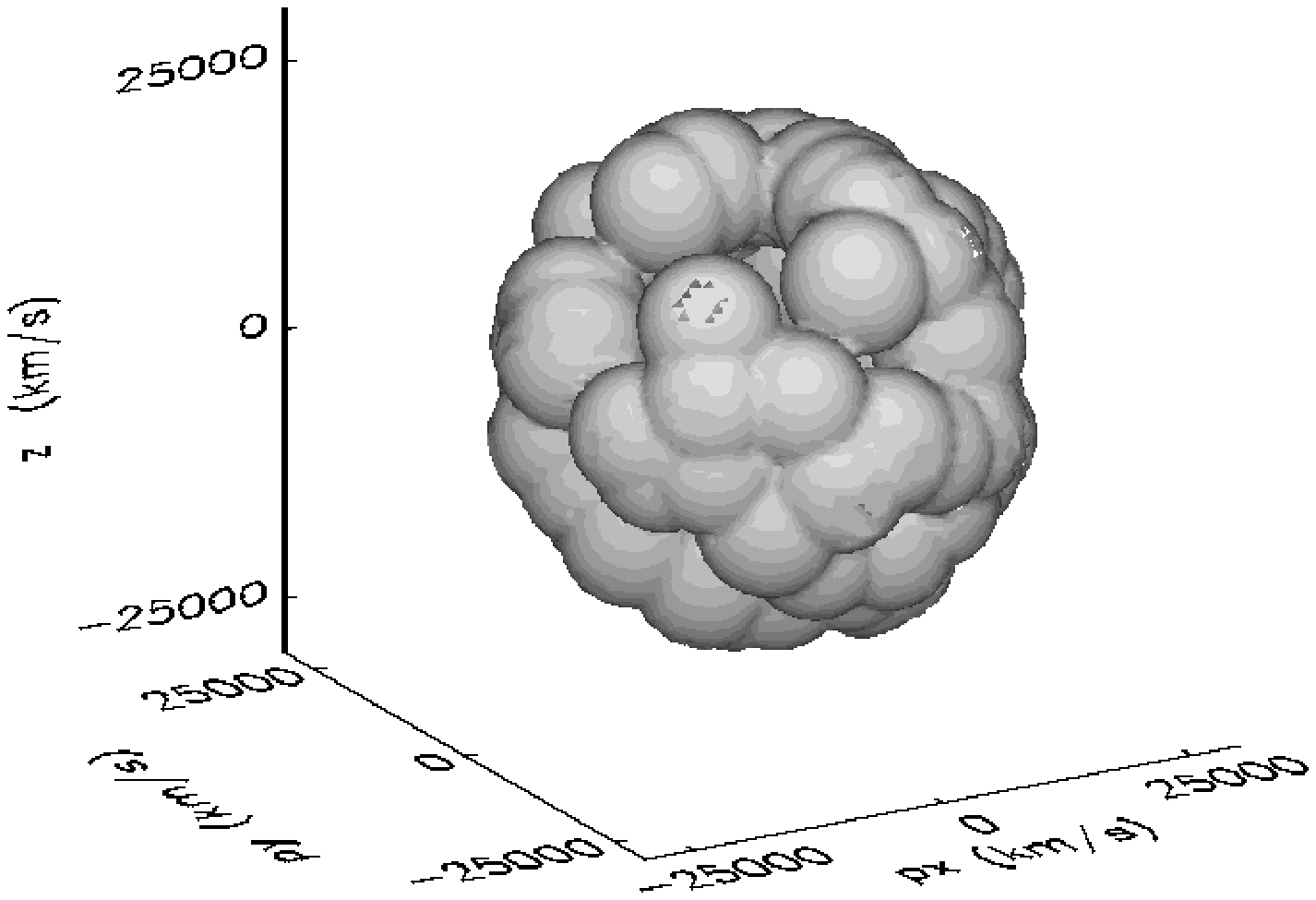]
{
 A sample clumpy model with 66\% of the envelope filled.
 \label{fig3}
}

\figcaption[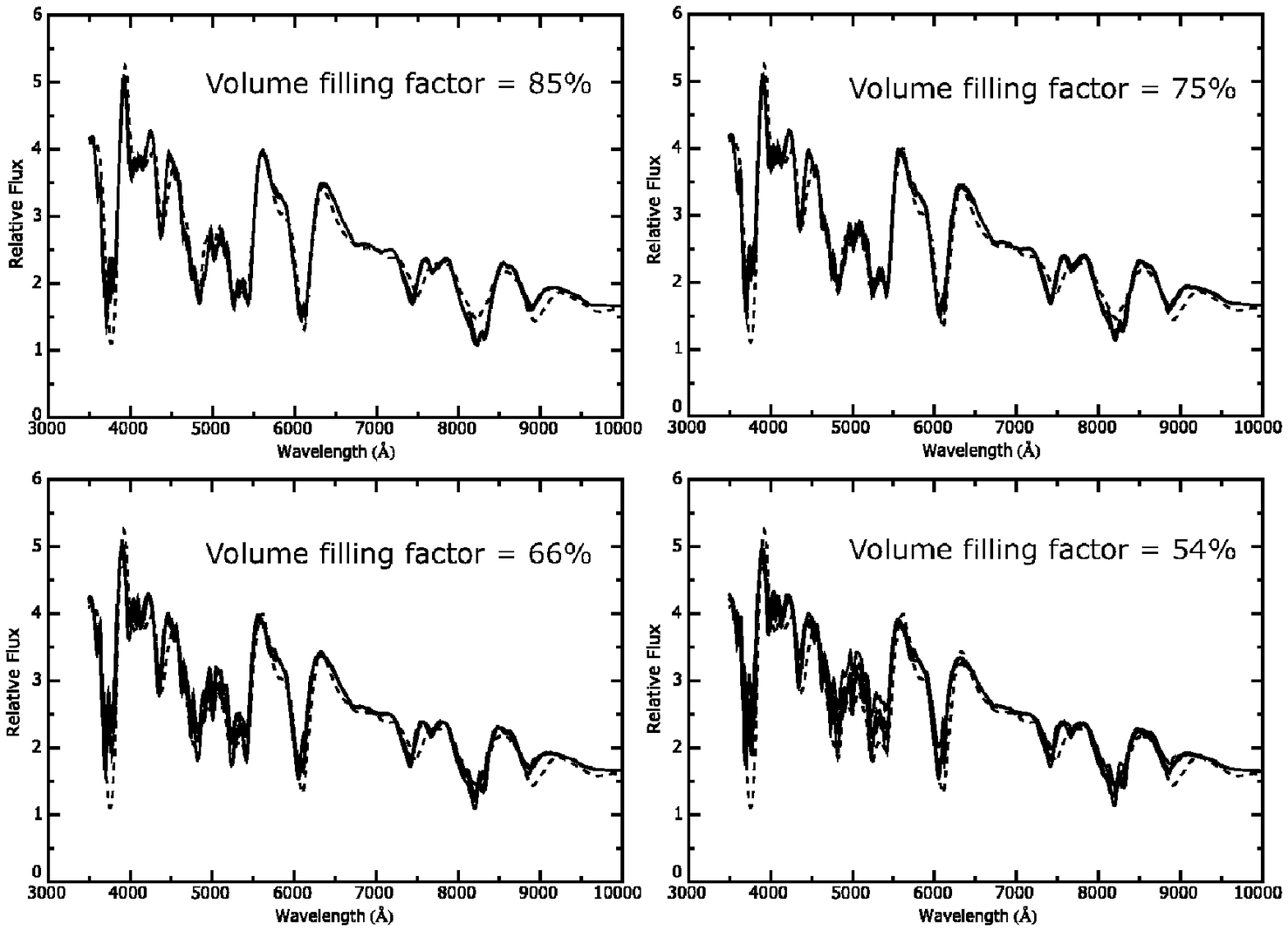]
{
 Spectra resulting from ClumpySYN calculations with different envelope
 filling factors (labelled in each panel).  A synthetic spectrum 
 similar to a maximum light fit to SN 1994D \citep{H99} is the dashed 
 line.
 \label{fig4}
}

\figcaption[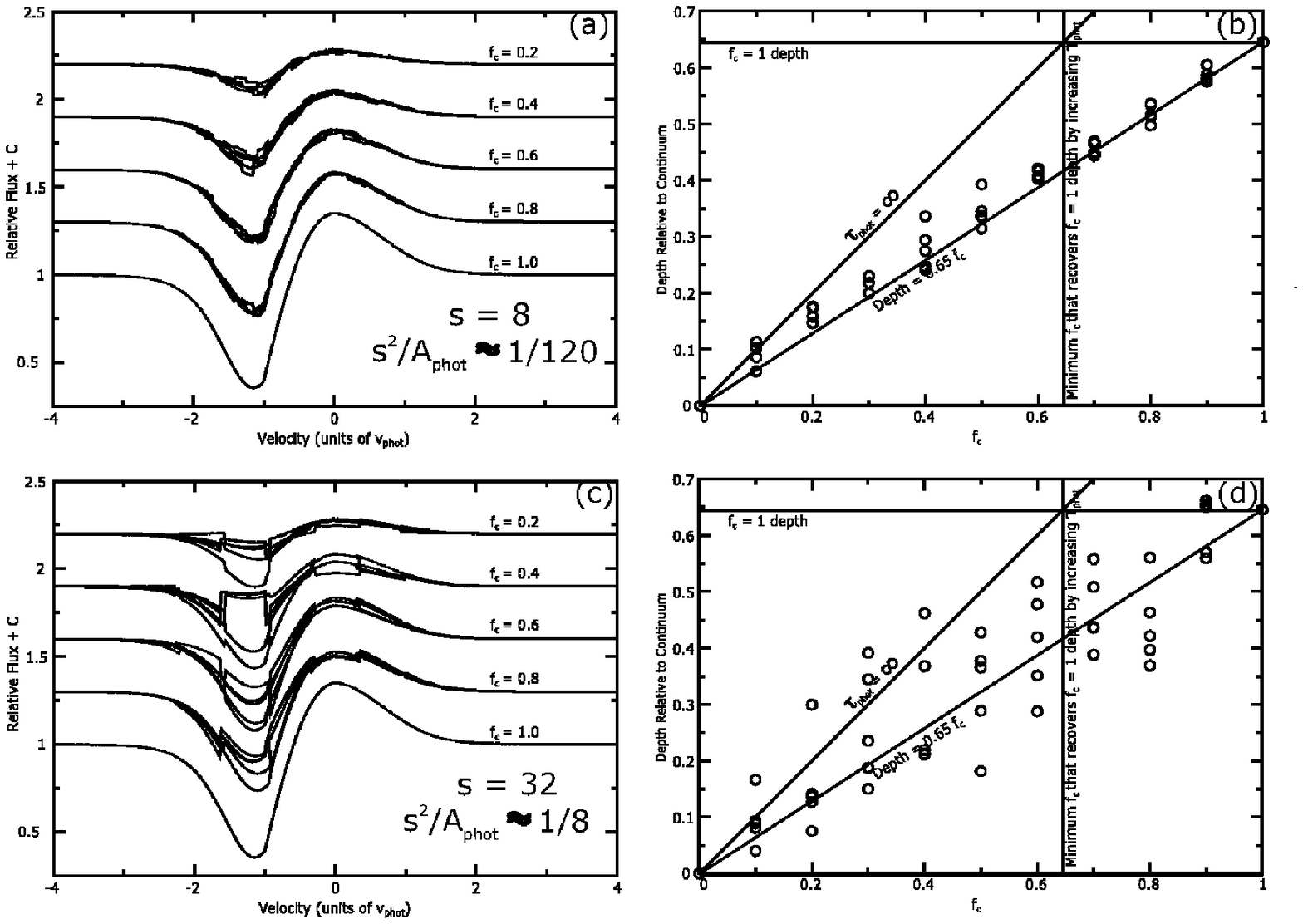]
{
 (a) Six line profiles each for $f_c = $ (0.2, 0.4, 0.6, 0.8 and 1.0)
 from top to bottom with small-area perturbations.  The lines are 
 offset by $C = $ (1.2, 0.9, 0.6, 0.3 and 0) from top to
 bottom.  
 (b) A corresponding plot of absorption depth as a function of $f_c$ for 
 the small perturbations.  Even if the line optical depth is increased
 to infinity, only for $f_c \gtrsim 0.7$ can the $f_c = 1$ depth be
 recovered.  
 (c) Same as (a) for large-area perturbations.  
 (d) The depth scatter compared to that for the small-area perturbations is
 much larger.  Increasing optical depth will on average recover the
 absorption depth for $f_c = 1$, but too much scatter in the
 absorption depth will be incurred.
 \label{fig5}
}

\figcaption[fig6.eps]
{
 Sample Type Ia Si II features at maximum light (about 20 d after explosion)
 for SN 1989B \cite{W94}, 
 SN 1990N \cite{M93}, SN 1994D \cite{P96} and SN 1998bu \cite{He00}.
}


\eject
\plotone{fig1.eps}
\eject
\plotone{fig2.eps}
\eject
\plotone{fig3.eps}
\eject
\plotone{fig4.eps}
\eject
\plotone{fig5.eps}
\eject
\plotone{fig6.eps}

\end{document}